\documentclass[cits]{PoS}
\usepackage{natbib,graphicx}

\newcommand{\thinsp}{\,}

\title{Supernova VLBI in the present and with the SKA}

\ShortTitle{Supernova VLBI}

\author{\speaker{Bietenholz, M. F.}\\
        York University, Toronto, ON, Canada, and\\
        Hartebeesthoek Radio Observatory, Krugersdorp, South Africa\\
        E-mail: \email{mbieten@yorku.ca}}

\abstract{VLBI is the only technology that will allow
sub-milliarcsecond resolution imaging in the near future.  As such, it
is the only way to image expanding supernovae in nearby galaxies.
Such images potentially allow us to study the early evolution of
neutron stars or black holes left behind by core-collapse supernovae,
the circumstellar wind history of the supernova progenitor stars, the
shock acceleration of cosmic-ray particles in supernovae as well as
the evolutionary process by which supernova shells merge into, and
enrich, the ISM\@.  I will discuss the results of the on-going VLBI
imaging campaigns on supernova 1986J and 1993J\@.  I will also discuss
the impact on supernova VLBI of the proposed South-African Karoo Array
Telescope and Australian ASKAP arrays, as well as the SKA itself, as
these telescopes will greatly increase the sensitivity of the global
VLBI network.}

\FullConference{From planets to dark energy: the modern radio universe\\
		 October 1-5 2007\\
		 University of Manchester, Manchester, UK}

\begin{document}
\section{Introduction}
Although there are many supernova remnants in our Galaxy, studying the
actual supernova events is difficult because the Galactic supernova
rate is only on the order of one per century, with the last Galactic
supernova occurring $\sim$300~yr ago.  Many more supernovae can be observed
in external galaxies, however, resolving them in the first few
decades of their lives requires milliarcsecond resolution.
Very-long-baseline interferometry (VLBI) is the only technology that
will allow sub-milliarcsec resolution imaging in the near future
(note that the diffraction limit of a 30~m optical telescope is
$\sim$8~milliarcsec at 10,000~\AA).

As the supernova ejecta plough out through the circumstellar medium
(CSM), a forward shock is driven into the CSM and a reverse shock is
driven into the ejecta.  These shocks, and associated instabilities,
can both amplify the magnetic field and accelerate particles to
relativistic energies, resulting in synchrotron radio emission.
Only supernovae with relatively dense CSM produce radio emission
detectable with current technology, and so far, only core-collapse
supernovae, that is, types Ib/c and II, have been studied with VLBI\@.

For the sample of supernovae which can be imaged with VLBI, the
rewards are rich.  VLBI allows us to study in detail the interaction
of the expanding ejecta with the CSM, which is in most cases, the
stellar wind from the supernova progenitor.  Over just a few years,
the supernova shock allows us to probe the last $\sim$10,000~yrs of
wind history of the progenitor.  VLBI imaging also allows us to study
the evolution of the supernova shell, the shock acceleration process, and
the possible emergence of a black hole or a neutron star.  VLBI
studies of supernovae are also an important tool for measuring the
supernova rate, and thus the star formation rate, in star-forming
regions, which are typically highly obscured in the optical.

Global VLBI at cm wavelengths can resolve young supernova shells out
to distances of 10 -- 20~Mpc. However, only a fraction of all
core-collapse supernovae are radio bright, and so far, only two have
been sufficiently close and radio bright to allow well resolved VLBI
images since shortly after the supernova explosion: SN~1986J (NGC~891,
$\sim$10~Mpc) and SN~1993J (M81, $\sim$4~Mpc), although several older
supernovae in M82 also have well-resolved VLBI images
\citep[e.g.,][]{Beswick+2006}.

I will discuss some of the current results from VLBI observations of
supernovae in the next few sections, and then proceed to elaborate on
the prospects of radio supernova imaging with the SKA demonstrators
and the SKA itself.

\section{SN 1993J}

\begin{figure}
\hspace{-0.16in}\includegraphics[angle=-90,width=1.10\textwidth]{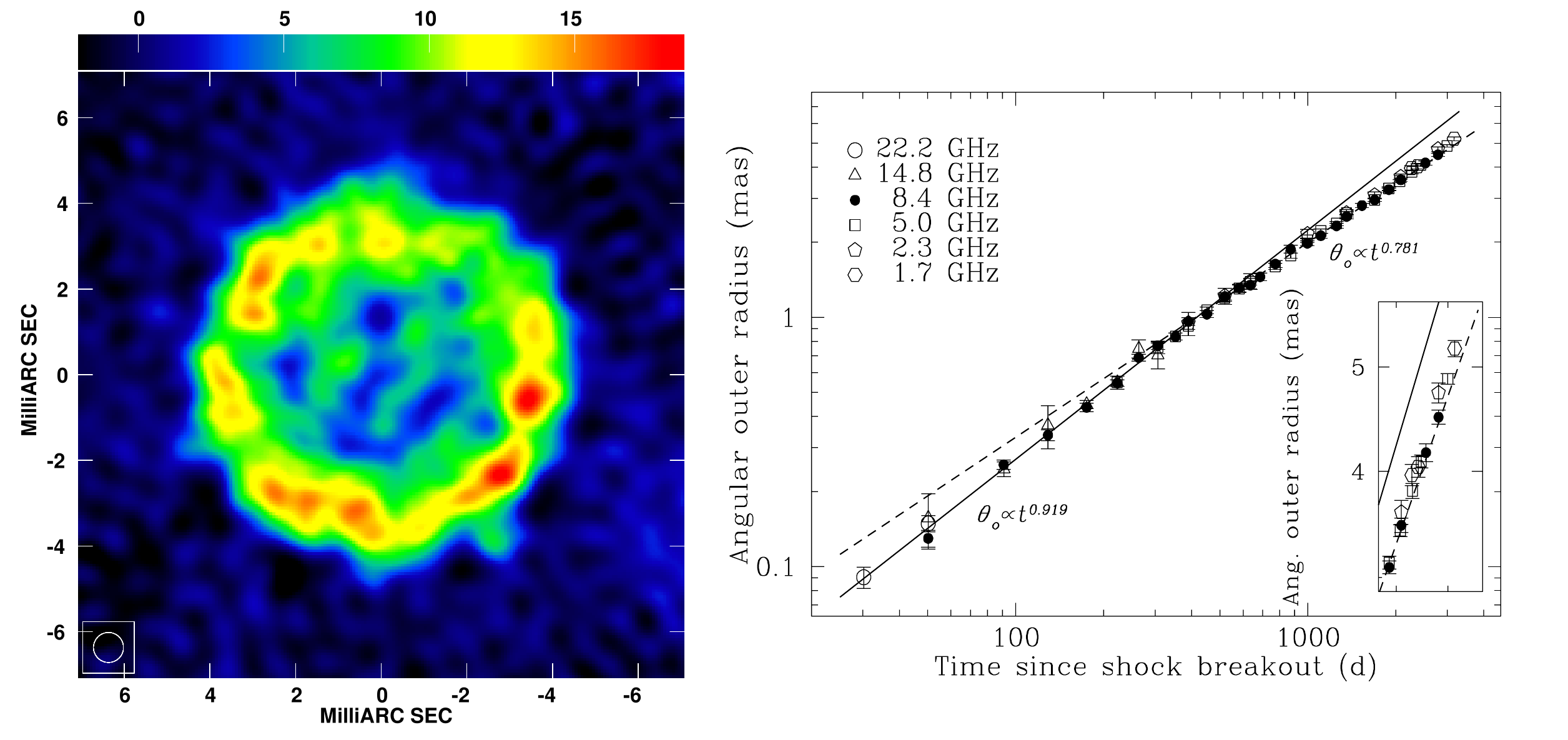}
\caption{Left panel: Composite image of SN~1993J at 8.4~GHz made by
combining the data sets from 1998 December, 2000 February and November
(at times, $t = 2080$~d, 2525~d and 2787~d).  The data sets were first
aligned and scaled to compensate for the expansion of the supernova,
\citep[for details see][]{SN93J-1}.  The
\protect\href{http://pos.sissa.it/archive/conferences/052/042/MRU_042_a1.mpg}
{\tt accompanying animation} shows the expansion of the supernova, and
was made from 27 epochs of VLBI observations (1993 to 2003) at 8.4 and
5~GHz \citep[for details, see][]{SN93J-3}. Right panel: the angular
outer radius, $\theta_{\rm o}$, of SN~1993J as a function of time,
$t$, showing the change of the exponent of the power-law expansion
near day 900, and with the inset showing the most recent measurements
\citep[for details, see][]{SN93J-2}.}
\label{fsn93jimage}
\end{figure}

The best studied radio supernova is SN~1993J\@.  It was one of the
brightest radio supernovae,
and high-quality radio light-curves were obtained over a wide range of
frequency \citep[e.g.,][]{SN93J-2, Weiler+2007}. The VLBI images showed a
remarkably symmetrical shell morphology
\citep[e.g.,][]{SN93J-Sci, Marcaide+1997}. The angular expansion
velocity could be accurately measured, and it is the first supernova
for which a changing deceleration rate could be measured
\citep[e.g.,][]{SN93J-2, Marcaide+2002b}.
Accurate astrometry has shown that the expansion of the shell is very
symmetrical about the explosion center \citep{SN93J-1, SN93J-3}.  
We show a high-resolution image of SN~1993J, accompanied by an
animation showing its expansion over a decade of VLBI imaging, in
Figure~\ref{fsn93jimage}.
By combining the angular expansion velocity measured with VLBI with
the radial expansion velocities determined from optical spectra, a
direct distance to M81 of $3.96 \pm 0.29$~Mpc could be determined
\citep{SN93J-4}.

\section{SN 1986J}

\begin{figure}
\centering
\includegraphics[width=3.3in]{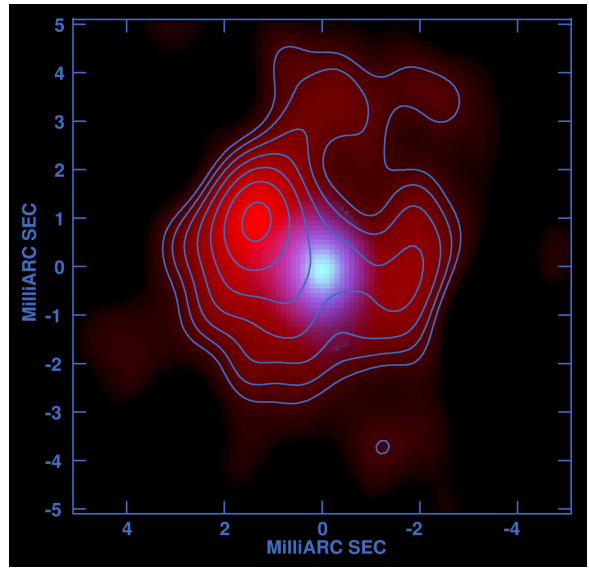}
\caption{A dual-frequency VLBI image of SN~1986J, showing the compact,
inverted-spectrum component located almost precisely in the center of
the expanding shell.  The red colour and the contours represent the
5~GHz radio brightness, showing the shell emission.  The contours are
drawn at 11.3, 16, 22.6, \dots, 90.5\% of the peak 5~GHz brightness of
0.55~mJy~beam$^{-1}$.  The blue through white represents the 15~GHz
radio brightness, showing the compact, central component which
appeared around 1999. North is up and east to the left. The
\protect\href{http://pos.sissa.it/archive/conferences/052/042/MRU_042_a2.mpg}
{\tt accompanying animation} shows the expansion of the supernova, and
the emergence of the central component.}
\label{fsn86jimage}
\end{figure}

Supernova 1986J was first discovered in the radio, a few years after
the explosion.  An early VLBI image of it marked the first time
shell-like structure was seen in an IAU-designated supernova
\citep{Bartel+1991}. Until 1999, SN~1986J had a power-law radio
spectrum ($S_\nu \propto \nu^\alpha$ where $\alpha \sim -0.6$) similar
to that of most radio supernovae.
After 1999, however, an inversion appeared in the spectrum.
Multi-frequency VLBI imaging showed that this inversion was associated
with a new component almost precisely in the center of the shell
\citep{SN86J-Sci}.  The new component is likely radio
emission associated with the black-hole or neutron-star compact
remnant of the explosion, which has not been seen in any other modern
supernova.  We show a two-frequency image of SN~1986J,
accompanied by an animation showing the evolution of the supernova
from almost 20~years of VLBI imaging, in Figure~\ref{fsn86jimage}.

\section{Other Radio Supernovae and Supernovae in Star-Forming Galaxies}

A number of other individual radio supernovae have been imaged with
VLBI: there is a series of VLBI observations of SN~1979C
\citep[e.g.,][]{SN79C}, and a handful of others including SN~2001em
\citep[e.g.,][the last being the first e-VLBI experiment on a
supernova]{SN2001em-2, Stockdale+2005, Paragi+2005}, SN~2001gd
\citep{Perez-Torres+2005} and SN~2004et \citep{Marti-Vidal+2007}.

Wide-field VLBI images of star-forming galaxies have yielded fruitful
results.  Wide-field VLBI observations of the nearby starburst M82
have shown a number of supernova and supernova remnants
\citep[e.g.,][]{Beswick+2006}.  The extreme starburst galaxy Arp~220
has been imaged with VLBI for over a decade, and the rate of new radio
supernovae is $4\pm2$~yr$^{-1}$ \citep{Conway+2007, Parra+2007}.
Radio observations of the starburst galaxy Arp~229 show that the radio
of radio supernovae must be about 1 per year \citep{NeffUT2004}.

\section{Supernova VLBI with the SKA Demonstrators}

Supernova VLBI is limited both by the available resolution and
sensitivity.  Although the SKA demonstrators and the SKA itself will
not allow an increase in resolution, they will greatly increase the
sensitivity as well as the $u$-$v$~coverage of the global VLBI array,
particularly in the southern hemisphere, where the network of
telescopes is notably less dense than in the northern hemisphere.
Presently, only the radio-brightest supernovae can be imaged: in
the last 30 years, only $\sim$10\% of the observed core-collapse
supernovae have been bright enough for a radio lightcurve to be
determined.

The increased sensitivity available, especially once the SKA is
operational, will also allow us detect far more supernovae and also to
follow individual supernovae for much longer.  It will allow us to
resolve older and more distant supernovae, for example Cas~A would
have a 1-GHz flux density of 1~$\mu$Jy and a size of 6~milliarcsec at
a distance of 170~Mpc.  We will be able to obtain a census of
supernova remnants of nearby galaxies, and much more routinely monitor
galaxies for the appearance of new radio supernova, which will do much
to constrain supernova and star formation rates.

\begin{table}
{\centering\small
\begin{tabular}{l c c c c c c}
\hline
   &                   &   +   &  +   &   +   &  +   &  +  \\
   & Australian LBA    & ASKAP & MeerKAT & ASKAP + &  SKA (core) & SKA (full) \\
   & \& Hartebeesthoek &       &         & MeerKAT  &  + ASKAP  & + ASKAP \\
   &                   &       &         &          &  +MeerKAT & + MeerKAT \\
\hline
Bandwith (MHz)    & 64  &  128 &  128 &  128 &  512 &  512 \\
Bitrate (Mbits/s) & 512 & 1024 & 1024 & 1024 & 4096 & 4096 \\
Image rms ($\mu$Jy) & 10 &  4.5 &  2.5 &  2.1 & 0.12 & 0.04 \\
\hline
\end{tabular}

\caption{Notes ---
All entries assume 8-hour observations using two polarizations
and 2-bit digitization;  ASKAP: assumes $45 \times 12$~m dishes with
system temperatures of 35~K; MeerKAT: assumes $80 \times 12$~m
dishes with system temperatures of 30~K; SKA (full): assumes a total
collecting area of $10^6$~m$^2$, distributed over 10 stations, with
25\% of collecting area at distances $>180$~km from the core.
}}
\end{table}

\newcommand{\araa}{Ann. Rev. Astron. Astrophys.}
\newcommand{\aap}{Astron. Astrophys.}
\newcommand{\aapr}{Astron. Astrophys. Rev.}
\newcommand{\aaps}{Astron. Astrophys. Suppl. Ser.}
\newcommand{\aj}{AJ}
\newcommand{\apj}{ApJ}
\newcommand{\apjl}{ApJL}
\newcommand{\apjs}{ApJS}
\newcommand{\apss}{ApSS}
\newcommand{\baas}{BAAS}
\newcommand{\memras}{Mem. R. Astron. Soc.}
\newcommand{\memsai}{Mem. Soc. Astron. Ital.}
\newcommand{\mnras}{MNRAS}
\newcommand{\iaucirc}{IAU Circ.}
\newcommand{\jrasc}{J. R. Astron. Soc. Can.}
\newcommand{\nat}{Nat}
\newcommand{\pasj}{PASJ}
\newcommand{\pasp}{PASP}


\begin{thebibliography}{99}

\bibitem{Beswick+2006}
R.~J. {Beswick}, J.~D. {Riley}, I.~{Marti-Vidal}, A.~{Pedlar}, T.~W.~B.
  {Muxlow}, A.~R. {McDonald}, K.~A. {Wills}, D.~{Fenech}, and M.~K. {Argo},
  {\it {15 years of very long baseline interferometry observations of two
  compact radio sources in Messier 82}},  {\em \mnras} {\bf 369} (July, 2006)
  1221--1228, [\href{http://xxx.lanl.gov/abs/arXiv:astro-ph/0603629}{{\tt
  arXiv:astro-ph/0603629}}].

\bibitem{SN93J-2}
N.~{Bartel}, M.~F. {Bietenholz}, M.~P. {Rupen}, A.~J. {Beasley}, D.~A.
  {Graham}, V.~I. {Altunin}, T.~{Venturi}, G.~{Umana}, W.~H. {Cannon}, and
  J.~E. {Conway}, {\it {SN 1993J VLBI. II. Related Changes of the Deceleration,
  Flux Density Decay, and Spectrum}},  {\em \apj} {\bf 581} (Dec., 2002)
  404--426.


\bibitem{Weiler+2007}
K.~W. {Weiler}, C.~L. {Williams}, N.~{Panagia}, C.~J. {Stockdale}, M.~T.
  {Kelley}, R.~A. {Sramek}, S.~D. {Van Dyk}, and J.~M. {Marcaide}, {\it {Long
  Term Radio Monitoring of SN 1993J}},  {\em ArXiv e-prints} {\bf 709} (Sept.,
  2007) [\href{http://xxx.lanl.gov/abs/0709.1136}{{\tt arXiv:0709.1136}}].

\bibitem{SN93J-Sci}
N.~{Bartel}, M.~F. {Bietenholz}, M.~P. {Rupen}, A.~J. {Beasley}, D.~A.
  {Graham}, V.~I. {Altunin}, T.~{Venturi}, G.~{Umana}, W.~H. {Cannon}, and
  J.~E. {Conway}, {\it {The Changing Morphology and Increasing Deceleration of
  Supernova 1993J in M81}},  {\em Science} {\bf 287} (Jan., 2000) 112--116.

\bibitem{Marcaide+1997}
J.~M. {Marcaide}, et al.,
 {\it {Deceleration in the Expansion of
  SN 1993J}},  {\em \apjl} {\bf 486} (Sept., 1997) L31.

\bibitem{Marcaide+2002b}
J.~M. {Marcaide}, et al.,
 {\it {How is really decelerating the
  expansion of SN1993J?}},  in {\em Proceedings of the 6th EVN Symposium},
  p.~239, June, 2002.

\bibitem{SN93J-1}
M.~F. {Bietenholz}, N.~{Bartel}, and M.~P. {Rupen}, {\it {SN 1993J VLBI. I. The
  Center of the Explosion and a Limit on Anisotropic Expansion}},  {\em \apj}
  {\bf 557} (Aug., 2001) 770--781. 
  [\href{http://xxx.lanl.gov/abs/arXiv:astro-ph/0104156}{{\tt
  arXiv:astro-ph/00104156}}].

\bibitem{SN93J-3}
M.~F. {Bietenholz}, N.~{Bartel}, and M.~P. {Rupen}, {\it {SN 1993J VLBI. III.
  The Evolution of the Radio Shell}},  {\em \apj} {\bf 597} (Nov., 2003)
  374--398.
  [\href{http://xxx.lanl.gov/abs/arXiv:astro-ph/0307382}{{\tt
  arXiv:astro-ph/0307382}}].

\bibitem{SN93J-4}
N.~{Bartel}, M.~F. {Bietenholz}, M.~P. {Rupen}, and V.~V. {Dwarkadas}, {\it {SN
  1993J VLBI. IV. A Geometric Distance to M81 with the Expanding Shock Front
  Method}},  {\em \apj} {\bf 668} (Oct., 2007) 924--940,
  [\href{http://xxx.lanl.gov/abs/arXiv:0707.0881}{{\tt arXiv:0707.0881}}].

\bibitem{Bartel+1991}
N.~{Bartel}, M.~P. {Rupen}, I.~I. {Shapiro}, R.~A. {Preston}, and A.~{Rius},
  {\it {A high-resolution radio image of a young supernova}},  {\em \nat} {\bf
  350} (Mar., 1991) 212--214.

\bibitem{SN86J-Sci}
M.~F. {Bietenholz}, N.~{Bartel}, and M.~P. {Rupen}, {\it {Discovery of a
  Compact Radio Component in the Center of Supernova 1986J}},  {\em Science}
  {\bf 304} (June, 2004) 1947--1949.

\bibitem{SN79C}
N.~{Bartel} and M.~F. {Bietenholz}, {\it {SN 1979C VLBI: 22 Years of Almost
  Free Expansion}},  {\em \apj} {\bf 591} (July, 2003) 301--315.

\bibitem{SN2001em-2}
M.~F. {Bietenholz} and N.~{Bartel}, {\it {SN 2001em: No Jet-driven Gamma-Ray
  Burst Event}},  {\em \apjl} {\bf 665} (Aug., 2007) L47--L50,
  [\href{http://xxx.lanl.gov/abs/arXiv:0706.3344}{{\tt arXiv:0706.3344}}].

\bibitem{Stockdale+2005}
C.~J. {Stockdale}, B.~{Kaster}, L.~O. {Sjouwerman}, M.~P. {Rupen},
  I.~{Marti-Vidal}, J.~M. {Marcaide}, S.~D. {van Dyk}, K.~W. {Weiler},
  B.~{Paczynski}, and N.~{Panagia}, {\it {Supernova 2001em in UGC 11794}},
  {\em \iaucirc} {\bf 8472} (Jan., 2005) 4.

\bibitem{Paragi+2005}
Z.~{Paragi}, M.~A. {Garrett}, B.~{Paczy{\'n}ski}, C.~{Kouveliotou},
  A.~{Szomoru}, C.~{Reynolds}, S.~M. {Parsley}, and T.~{Ghosh}, {\it {e-VLBI
  observations of SN2001em - an off-axis GRB candidate .}},  {\em Memorie della
  Societa Astronomica Italiana} {\bf 76} (2005) 570,
  [\href{http://xxx.lanl.gov/abs/arXiv:astro-ph/0505468}{{\tt
  arXiv:astro-ph/0505468}}].

\bibitem{Perez-Torres+2005}
M.~A. {P{\'e}rez-Torres} et al.,
  {\it {High-resolution observations of SN
  2001gd in NGC 5033}},  {\em \mnras} {\bf 360} (July, 2005) 1055--1062,
  [\href{http://xxx.lanl.gov/abs/arXiv:astro-ph/0504647}{{\tt
  arXiv:astro-ph/0504647}}].

\bibitem{Marti-Vidal+2007}
I.~{Mart{\'{\i}}-Vidal} et al., 
  {\it {8.4 GHz VLBI observations of
  SN{\thinsp}2004et in NGC{\thinsp}6946}},  {\em \aap} {\bf 470} (Aug., 2007)
  1071--1077, [\href{http://xxx.lanl.gov/abs/arXiv:0705.3853}{{\tt
  arXiv:0705.3853}}].

\bibitem{Conway+2007}
J.~E.~{Conway}, {\em these proceedings}, \pos{PoS(MRU)045}.

\bibitem{Parra+2007}
R.~{Parra}, J.~E. {Conway}, P.~J. {Diamond}, H.~{Thrall}, C.~J. {Lonsdale},
  C.~J. {Lonsdale}, and H.~E. {Smith}, {\it {The Radio Spectra of the Compact
  Sources in Arp 220: A Mixed Population of Supernovae and Supernova
  Remnants}},  {\em \apj} {\bf 659} (Apr., 2007) 314--330,
  [\href{http://xxx.lanl.gov/abs/arXiv:astro-ph/0612248}{{\tt
  arXiv:astro-ph/0612248}}].

\bibitem{NeffUT2004} 
S.~G.~{Neff}, J.~S. {Ulvestad}, and S.~H. {Teng}, {\it {A Supernova
  Factory in the Merger System Arp 299}}, {\em \apj} {\bf 611} (Aug.,
  2004) 186--199,
  [\href{http://xxx.lanl.gov/abs/arXiv:astro-ph/0406421}{{\tt
  arXiv:astro-ph/0406421}}]

\end{thebibliography}
\end{document}